\documentclass[%
 reprint,
 amsmath,amssymb,
 aps,
]{revtex4-2}

\usepackage{graphicx}
\usepackage{dcolumn}
\usepackage{bm}

\begin{document}

\preprint{APS/123-QED}

\title{The vertical-velocity skewness in the inertial sublayer of turbulent wall flows}

\author{Elia Buono$^{1,2*}$}
\author{Gabriel Katul$^{1}$} 
\author{Michael Heisel$^{3}$}
\author{Davide Vettori$^2$}
\author{Davide Poggi$^{2}$}
\author{Cosimo Peruzzi$^{4}$}
\author{Costantino Manes$^{2}$}

\affiliation{$^1$Department of Civil and Environmental Engineering, Duke University, Durham, NC, USA}

\affiliation{$^2$Dipartimento di Ingegneria dell'Ambiente, del Territorio e delle Infrastrutture, Politecnico di Torino, Torino, Italia}

\affiliation{$^3$School of Civil Engineering, University of Sydney, Sydney, Australia}

\affiliation{$^4$Area for Hydrology, Hydrodynamics, Hydromorphology and Freshwater Ecology, Italian Institute for Environmental Protection and Research, Rome, Italy}

\email{Elia Buono: elia.buono@polito.it}

\date{\today}
\begin{abstract}
We provide empirical evidence that within the inertial sub layer of adiabatic turbulent flows over smooth walls, the skewness of the vertical velocity component $Sk_w$ displays universal behaviour, being constant and constrained within the range $Sk_w \approx 0.1-0.16$, regardless of flow configuration and Reynolds number. A theoretical model is proposed to explain the observed behaviour, including the observed range of variations of $Sk_w$. The model clarifies why $Sk_w$ cannot be predicted from down-gradient closure approximations routinely employed in meteorological and climate models whereby $Sk_w$ impacts cloud formation and dispersion processes. The model also offers an alternative and implementable approach.
\end{abstract}

z\maketitle

\section{Introduction}
\label{sec:introduction}
Much of the effort devoted to the study of adiabatic and hydrodynamically smooth wall turbulence has focused on the characterization of velocity statistics within the so-called logarithmic or inertial sublayer (ISL). The Attached Eddy Model (AEM), which is probably the most cited model for ISL-turbulence, predicts that first and second order velocity statistics can be described as \citep{townsend1980structure,smits2011high, marusic2019attached}: 
\begin{equation}
\label{eqn:law_of_wall}
\overline{u}^+ = \frac{1}{\kappa}\log\left({z^+}\right)+A;~ 
\sigma_u^{2+} = A_u-B_u \log\left(\frac{z}{\delta}\right);
\end{equation}
and, a less studied outcome, $\sigma_w^{2+} = A_w^2$, where $u$ and $w$ are the longitudinal and wall-normal velocity component, respectively; $z$ is the wall normal coordinate; $\sigma_u=\sqrt{\overline{u'^2}}$ and $\sigma_w=\sqrt{\overline{w'^2}}$ are the standard deviation of $u$ and $w$ respectively; primes identify fluctuations due to turbulence around the mean; the overline represents averaging over coordinates of statistical homogeneity; the plus index indicates classical inner scaling whereby velocities and lengths are normalized with the friction velocity $u_*$ and viscous length scale $\nu/u_*$, respectively, with $\nu$ being the kinematic viscosity of the fluid; $\delta$ is the outer length scale of the flow; $\kappa$, $A$, $A_u$, $A_w$, $B_u$ are coefficients that are thought to attain asymptotic constant values at very large Reynolds numbers $Re_{\tau}=u_*\delta/\nu$ \citep{marusic2013logarithmic, smits2011high,stevens2014large}.  The AEM has been extended to velocity moments of any order as well as cross-correlations between different velocity components thereby providing an expanded picture of ISL flow statistics \citep{woodcock2015statistical}.  However, convincing empirical support for the aforementioned theoretical predictions is limited to the statistics of $u$ \citep{smits2011high, marusic2013logarithmic,meneveau2013generalized,banerjee2013logarithmic, huang2022profiles}. In contrast, the statistics of $w$ have been much less reported and investigated, partly because of the technical difficulties associated with accurately measuring $w$ in the near wall region of laboratory flows at high $Re_{\tau}$. As a result, theoretical predictions of $w-$statistics have received mixed support from the literature \citep{morrill2015temporally,orlu2017reynolds,zhao2007scaling} and higher order moments of $w'$ are rarely reported but with few notable exceptions \citep{flack2007examination,schultz2007rough,manes2011turbulent,peruzzi2020scaling,heisel2020velocity}.  We argue that this overlook contributed to hide a universal property of ISL turbulence, which is herein reported and discussed. The aim of this letter is to demonstrate that the skewness of $w'$, $Sk_w=\overline{w'^3} /\sigma_w^3$, is constant (i.e. $z-$independent) and robust to variations in $Re_\tau$ within the ISL. Moreover, a theoretical model that explains this observed behaviour and links $Sk_w$ to established turbulence constants is proposed, leading to satisfactory predictions. Finally, we note that the asymmetry in the probability density function of $w'$, as quantified by $Sk_w$, cannot be accounted for with gradient-diffusion representations routinely employed in meteorological and climate models \citep{mellor1982development}. Rectifying this limitation is of significance because $Sk_w$ was shown to impact cloud formation \citep{bogenschutz2012unified,huang2020assessing,li2022updated} and dispersion processes \citep{baerentsen1984monte,luhar1989random,wyngaard1991transport,maurizi1999velocity}.

\begin{figure}
\centering
\noindent\includegraphics[width=8.7cm]{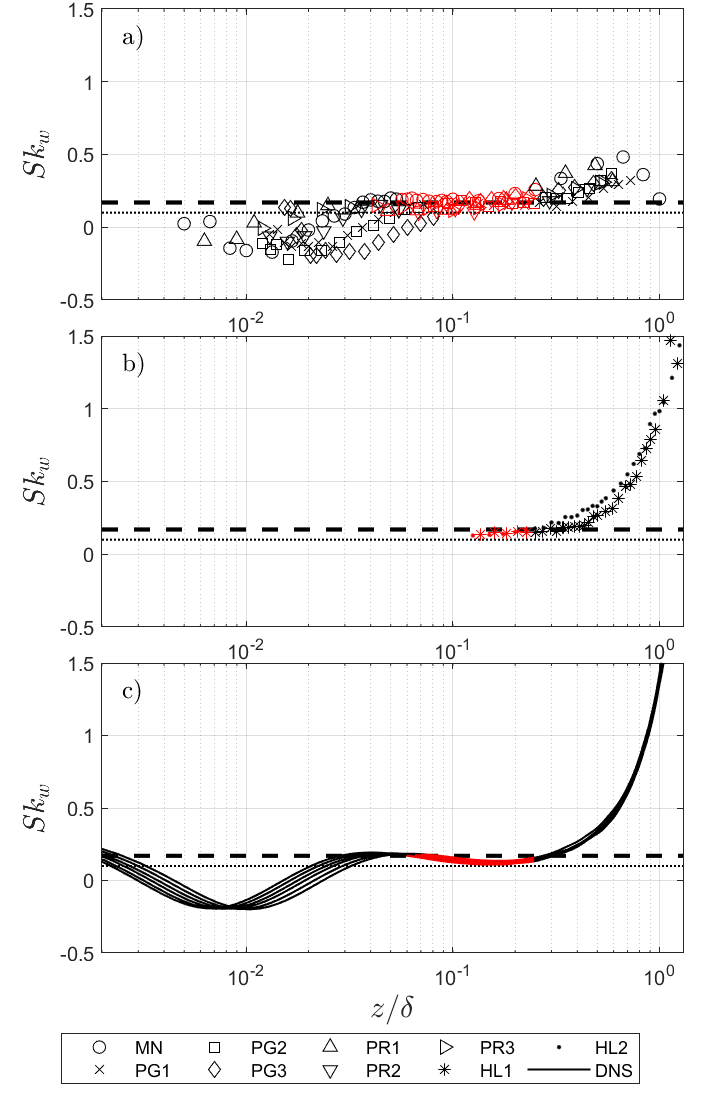}
\caption{Variation of the vertical velocity skewness $Sk_w$ with normalized wall-normal distance $z/\delta$. The dashed line is $Sk_w=0.16$ and the dotted line is $Sk_w=0.10$. Data sources and references are summarized in Table \ref{tab:table1}. Red symbols and lines identify the ISL range. For HL1 and HL2, near wall data are not reported due to issues with velocity measurements associated with an excessive sampling volume of the x-probe employed in the experiments \citep{heisel2020velocity}.}
\label{Fig:Skw_profile}     
\end{figure}

Figure \ref{Fig:Skw_profile} reports the variations of $Sk_w$ with normalized wall-normal distance ($z/\delta$) using data from Direct Numerical Simulations (DNS) \citep{sillero2013one} and laboratory experiments \citep{heisel2020velocity} pertaining to flat plate turbulent boundary layers (TBLs). Data from experiments on uniform \citep{poggi2002experimental} and weakly non-uniform turbulent open channel flows \citep{manes2011turbulent,peruzzi2020scaling}, whereby accurate measurements of $w$ are available are also included.  This data refers to flows within the low to medium range of $Re_{\tau}$ (Table \ref{tab:table1}). A reference value of $Sk_w=0.1$ is added to the figure as often reported for atmospheric surface layers in adiabatic conditions across multiple heights and surface cover \citep{chiba1978stability}. This value is representative of flows at extremely high $Re_{\tau}$. A region of constant $Sk_w$ weakly varying between 0.1 and 0.16 (here weakly means that variations are much smaller than those displayed by $Sk_w$ over the entire flow domain) is evident in all profiles within the range $2.6\sqrt{Re_\tau}\nu/u_*$ up to $0.15-0.25\delta$, which is often associated with the ISL \citep{zhou2015properties,orlu2016high,orlu2017reynolds}. This finding is remarkable given the large differences in $Re_{\tau}$, measurement techniques and experimental facilities used. In what follows, a theoretical model that predicts and explains such a behaviour is provided.  
\section{Theory}
\label{sec:theory}
To explain the observed behaviour of $Sk_w$, a stationary and planar homogeneous incompressible flow in the absence of subsidence is considered for $\overline{w'^3}$.  For these conditions, the model can be derived from the Reynolds averaged Navier-Stokes equations and is given as \citep{canuto1994second,zeman1976modeling}
\begin{align}
\label{eq:budgetwww1}
\frac{\partial \overline{w'^3}}{\partial t}=0&=+\overbrace{3 \sigma_w^2\frac{\partial \sigma_w^2}{\partial z}}^{{\rm Source/Sink}}-\overbrace{\frac{\partial \overline{w'w'^3}}{\partial z}}^{{\rm Turbulent~transport}}\\ &-\underbrace{3\left( \overline{w'w' \frac{\partial p'}{\partial z}}\right)}_{{\rm Pressure~velocity~destruction}} -\underbrace{2\nu \left( 3 \overline{w' \frac{\partial w'}{\partial z} \frac{\partial w'}{\partial z}}\right)}_{{\rm Viscous~destruction}},\nonumber
\end{align} 
where $t$ is time and $p'$ is the pressure deviation from the mean or hydrostatic state normalized by a constant fluid density $\rho$.  

The first two terms on the right-hand side of equation \ref{eq:budgetwww1} arise from inertial effects or convective acceleration, the third and fourth terms arise due to interactions between $w'w'$ and the forces acting on a fluid element ($p'/\rho$ and viscous stresses). A quasi-normal approximation for the fourth moment is used \citep{andre1976turbulence} so that $FF_w=\overline{w'^4}/(\sigma_w)^4=3+a$ and the overall inertial term simplifies to
\begin{equation}
\label{eq:budgetwww2}
-\frac{\partial \overline{w'^4}}{\partial z}+3 \sigma_w^2\frac{\partial \sigma_w^2}{\partial z}=-(3+2 a) \sigma_w^2 \frac{\partial\sigma_w^2}{\partial z},
\end{equation} 
where $a\neq0$ allows for deviations from Gaussian tails ($a=0$ recovers a Gaussian flatness factor).  Usage of a quasi-Gaussian approximation to close a fourth (and even) moment budget makes no statement on the asymmetry (or odd moments) of the $w'$ probability density function, only that large-scale intermittency is near-Gaussian, a finding well supported in the literature \citep{meneveau1991analysis} and many phenomenological approaches \citep{woodcock2015statistical}. Models for the pressure velocity and viscous destruction terms are now needed to integrate equation \ref{eq:budgetwww1}. A return to isotropy (or Rotta) model \citep{Rotta1951} given by
\begin{align}
\label{eq:4}
-2\overline{w'\frac{\partial p'}{\partial z}}=\frac{C_R}{\tau}\left(\frac{\overline{q}}{3} -\sigma_w^2 \right),
\end{align}
may be used to derive an expression for the pressure-velocity destruction term in in equation \ref{eq:budgetwww1} where $q=u'u'+v'v'+w'w'$ is twice the instantaneous turbulent kinetic energy, $\overline{q}=2K$, $K$ is the averaged turbulent kinetic energy, $v'$ is the lateral turbulent velocity, and $C_R=1.8$ is a well established constant, the Rotta constant \citep{bou2018role}.  The $C_R$ relates the so-called relaxation time $\tau=\overline{q}/\overline{\epsilon}$ to the time it takes for isotropy to be attained at the finest scales, where $\overline{\epsilon}$ is the mean turbulent kinetic energy dissipation rate.  In deriving such an extension, the non-averaged form of equation \ref{eq:4} is first multiplied by $(3/2)w'$ and then averaged to yield 
\begin{align}
\label{eq:5}
-3\left( \overline{w'w' \frac{\partial p'}{\partial z}}\right)=\frac{3}{2}\frac{C_R}{\tau_s}\left(\frac{\overline{w'q}}{3} -\overline{w'w'^2} \right),
\end{align}
where $\tau_s$ is another decorrelation time that can differ from $\tau$.  The difference between $\tau$ and $\tau_s$ arises because when multiplying the instantaneous form of equation \ref{eq:4} by $w'$ to arrive at the instantaneous form of equation \ref{eq:5}, the correlation between $q$ and $w'$ must be considered after averaging.  While expected to be small relative to the pressure-velocity interaction term, the viscous destruction contribution is herein retained and represented as \citep{zeman1976modeling} 
\begin{equation}
-2 \nu \left( 3 \overline{w' \frac{\partial w'}{\partial z} \frac{\partial w'}{\partial z}}\right)=-2 \overline {\epsilon' w'}= -c_2 \frac{\overline{w'q}}{\tau_s},
\end{equation}
where $c_2$ is a similarity constant, and $\epsilon'\sim q/\tau_s$ is the fluctuating dissipation rate around $\overline{\epsilon}$.  Inserting these approximations into equation \ref{eq:budgetwww1} yields, 
\begin{equation}
\label{eq:budgetwww2}
\overline{w'^3}=-\frac{2}{3}\frac{(3+2 a)\tau_s \sigma_w^2}{C_R}\frac{\partial \sigma_w^2}{\partial z}+\overline{w'q} \left( \frac{1}{3} - \frac{2 c_2}{3 C_R}\right).
\end{equation} 
\begin{table}[b]
\caption{\label{tab:table1}
Overview of smooth-wall boundary layer experiments (OC = open channel/flumes, WT = wind tunnel) and direct numerical simulation* in Figure \ref{Fig:Skw_profile}. The $Re_{\tau}=\delta u_*/\nu$ is the friction Reynolds number, $B_u$ and $A_w$ were computed from data using AEM.  For the DNS, the highest and lowest $Re_{\tau}$ are shown given the small variability in $B_u$ (0.91-0.94) and $A_w$ (1.15-1.17).  The computed $Sk_w$ using equation \ref{eq:budgetwww6} is also presented.}
\begin{ruledtabular}
\begin{tabular}{cccccccc}
               Source & Data set & Flow & $Re_{\tau}$ & $B_u$ & $A_w$ & $Sk_w$  \\ \hline
          \citet{manes2011turbulent} & MN    & OC & 2160 & 0.58 & 1.06 & 0.11 \\ \hline
           \citet{sillero2013one}    & DNS*  &    & 1307 & 0.85 & 1.15 & 0.13 \\
                                     &       &    & 2000 & 0.86 & 1.17 & 0.12 \\ \hline
         \citet{heisel2020velocity}  & HL1   & WT & 3800 & 0.85 & 0.96 & 0.21  \\ 
                                   ~ & HL2   & WT & 4700 & 0.63 & 1.00 & 0.15 \\ \hline  
       \citet{poggi2002experimental} & PG1   & OC & 1232 & 0.73 & 0.90 & 0.23 \\ 
                                   ~ & PG2   & OC & 1071 & 0.78 & 1.02 & 0.17 \\ 
                                   ~ & PG3   & OC & 845  & 1.03 & 0.90 & 0.33\\ \hline
        \citet{peruzzi2020scaling}   & PR1   & OC & 2240 & 0.78 & 1.12 & 0.13 \\ 
                                   ~ & PR2   & OC & 999  & 0.63 & 1.06 & 0.12\\ 
                                   ~ & PR3   & OC & 1886 & 0.85 & 1.06 & 0.16 \\ 
        
    \end{tabular}
    \end{ruledtabular}
\end{table}
A model for $\overline{w'q}$ is further needed to infer $Sk_w$.  To arrive at this model, the $K$ budget for the same flow conditions leading to equation \ref{eq:budgetwww1} are employed. When mechanical production is balanced by $\overline{\epsilon}$ as common to the ISL, the $K$ budget leads to two outcomes \citep{lopez1999wall}
\begin{equation}
\label{eq:tke}
u_*^2\frac{\partial \overline{u}}{\partial z}-\overline{\epsilon}=0; ~~-\frac{1}{2}\frac{\partial \overline{w'q}}{\partial z}=0.
\end{equation} 
The height-independence of $\overline{w'q}$ is suggestive that it must be controlled by local conditions and a down-gradient approximation is justified given by \citep{lopez1999wall}
\begin{equation}
\label{eq:budgetwww3}
-\frac{1}{2}\overline{w'q}=\kappa z u_* \frac{\partial K}{\partial z}.
\end{equation} 
The model in equation \ref{eq:budgetwww3} has received experimental support even for rough-wall turbulent boundary layers and across a wide range of Reynolds numbers and surface roughness values \citep{lopez1999wall}.  Noting that $K \approx \sigma_u^2$ yields
\begin{align}
\label{eq:budgetwww5}
\overline{w'^3} =&-\frac{2}{3}\left[K_{t,w}\frac{\partial \sigma_w^2}{\partial z}+  K_{t,u}\frac{\partial \sigma_u^2}{\partial z}\right];\\ K_{t,w}=&\frac{(3+2 a)\tau_s \sigma_w^2}{C_R}; ~ K_{t,u}=\kappa z u_* \left( 1 - \frac{2 c_2}{C_R}\right),\nonumber
\end{align} 
where $K_{t,w}$ and $K_{t,u}$ are eddy viscosity terms. These two eddy viscosity values become comparable in magnitude when setting $\tau_s=\kappa z/u_*$ (i.e. following classical ISL scaling) and $C_R=1.8$ - its accepted value \citep{bou2018role} as expected in the ISL.  To determine $\partial{\sigma_w^2}/{\partial z}$, the mean vertical velocity equation is considered for the same idealized flow conditions as equation \ref{eq:budgetwww1}.  This consideration results in  
${\partial \sigma_w^2}/{\partial z}=-({1}/{\rho})({\partial \overline{P}}/{\partial z})-g$, where $g$ is the gravitational acceleration. When $\overline{P}=-\rho g z$ (i.e. hydrostatic), $\partial\sigma_w^2/\partial z=0$ or $A_w$ is constant in $z$ within the ISL.  That is, the AEM requires $\overline{P}$ to be hydrostatic. However, the AEM precludes a $\partial \sigma_u^2/\partial z=0$ in the ISL.  In fact, the AEM predicts a $\partial \sigma_u^2/\partial z=-u_*^2 B_u/z$ when the $Re_{\tau}$ is very large as expected in the ISL of an adiabatic atmosphere.  Inserting this estimate into equation \ref{eq:budgetwww5}, setting $u_*=\sigma_w/A_w$ and momentarily ignoring $\partial{\sigma_w^2}/{\partial z}$ relative to $\partial \sigma_u^2/\partial z$ as a simplification consistent with the AEM, leads to
\begin{equation}
\label{eq:budgetwww6}
Sk_w=\frac{\overline{w'^3}}{\sigma_w^3}=\frac{2}{3} \left( 1 - \frac{2 c_2}{C_R}\right) \frac{\kappa B_u}{A_w^3}.
\end{equation} 
This equation is the sought outcome. The term $2 c_2/C_R$ reflects the relative importance of the pressure-velocity to viscous destruction terms. Pressure-velocity destruction effects are far more efficient than viscous effects supporting the argument that $2 c_2/C_R<<1$ at very high $Re_{\tau}$ \citep{katul2013co} such as the atmosphere.  This implies that the numerical value of $Sk_w$, as obtained from equation \ref{eq:budgetwww6}, depends on three well established phenomenological constants, namely $\kappa$, $A_w$, and $B_u$ \citep{banerjee2013logarithmic,marusic2019attached,huang2022profiles}, which, in turn, may depend weakly on $Re_{\tau}$ and the flow type.  Equation \ref{eq:budgetwww6} is also insensitive to the choices made for $\tau_s$, because the AEM requires $\partial\sigma_w^2/\partial z=0$.

\section{Discussion and Conclusion}
\label{sec:conclusion}

\begin{figure}
\centering
\noindent\includegraphics[width=8.7cm]{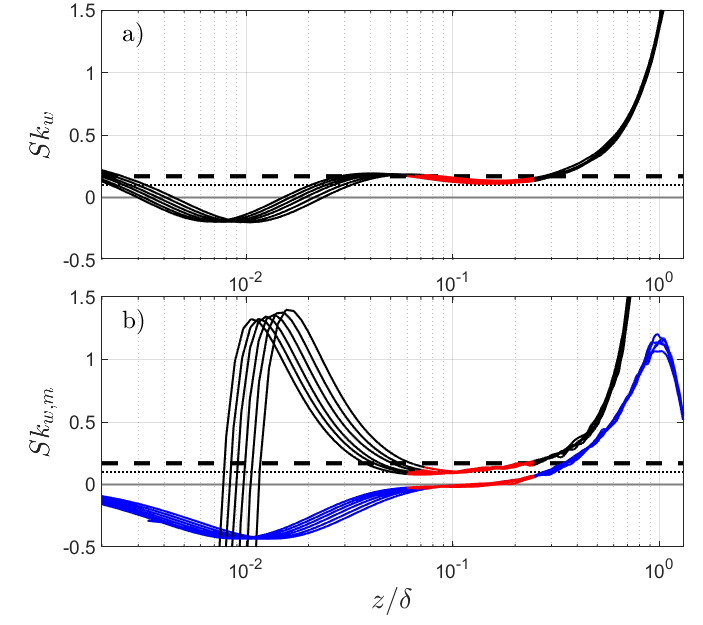}
\caption{Panel a): Variation of the vertical velocity skewness $Sk_w$ with normalized wall-normal distance $z/\delta$ from DNS \cite{sillero2013one}; panel b): $Sk_{w,m}$ is the modelled vertical velocity skewness using the first term (blue line) and second term (black line) on the right hand side of equation \ref{eq:budgetwww5} both scaled with $\sigma_w^3$.  In both panels, red lines identify the ISL range. The dashed line is $Sk_w=0.16$ and the dotted line is $Sk_w=0.10$}
\label{Fig:DNS_profile}     
\end{figure}

\begin{figure}
\centering
\noindent\includegraphics[width=8.8cm]{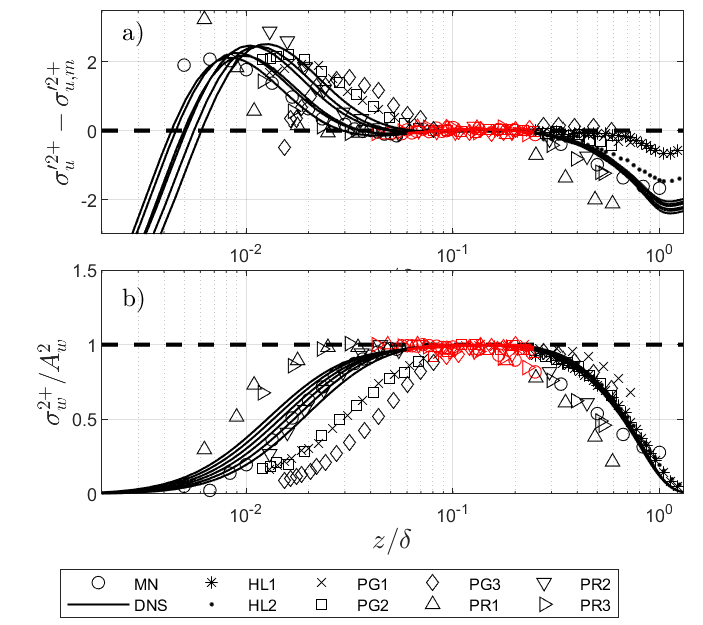}
\caption{Panel a): difference between  $\sigma_u^{2+}$ and estimations obtained from the AEM, $\sigma_{u,m}^{2+}=A_u-B_u \log(z/\delta)$ using values of $A_u$ and $B_u$ obtained from regression of data within the ISL range (identified by red symbols and lines); panel b): non dimensional vertical velocity variance $\sigma_w^2$ normalized with $A_w$ obtained from data fitting within the ISL vs wall-normal distance $z/\delta$. Data sources and references are summarized in Table \ref{tab:table1}.}
\label{Fig:AEM_profile}     
\end{figure}

Equation \ref{eq:budgetwww6} demonstrates two inter-related aspects about $Sk_w$ in the ISL: (i) why $Sk_w$ is positive and constant with $z$, and (ii) why conventional gradient diffusion approximations fail to predict $\overline{w'^3}$ from $\partial \sigma_w^2/\partial z$.  Regarding the first, equation \ref{eq:budgetwww6} predicts that $Sk_w>0$ consistent with the paradigm that ejective eddy motions ($w'>0, u'<0$) are more significant in momentum transfer than sweeping motions ($w'<0, u'>0$)  within the ISL. This assertion is supported by numerous experiments and simulations \citep{nakagawa1977prediction,raupach1981conditional,heisel2020velocity} and adds further confidence in the physics associated with the derivation of equation \ref{eq:budgetwww6}. Moreover, values of the constants in equation \ref{eq:budgetwww6} for flat plate turbulent boundary layers (TBL) at $Re_{\tau}\to\infty$ correspond to $\kappa=0.39$, $A_w=1.33$, and $B_u=1.26$ \citep{smits2011high,huang2022profiles}. Upon further setting $c_2=0.1$ and $C_R=1.8$ (conventional values), leads to $Sk_w=0.12$. This estimate compares well with $Sk_w=0.1$ reported for the ISL in the adiabatic atmosphere \citep{chiba1978stability,barskov2023relationships}. At the low to moderate $Re_{\tau}$ pertaining to the experiments and DNS associated with Figure \ref{Fig:Skw_profile}, equation \ref{eq:budgetwww6} cannot be used to estimate $Sk_w$ using asymptotic values of $A_w$ and $B_u$. However, Figure \ref{Fig:Skw_profile} shows that these flows attain similar (i.e. slightly higher) and reasonably z-independent values of $Sk_w$. To explain this behaviour, the DNS data are used as they allow exploring theoretical predictions offered by equation \ref{eq:budgetwww5} with reliable estimations of $\sigma_w^2$ and $\sigma_u^2$ vertical gradients. Figure \ref{Fig:DNS_profile} indicates that, for mostof the ISL, the first term on the right hand side of the equation is an order of magnitude smaller than the second and can be discarded as predicted by the AEM and advocated in the proposed theory. Predictions of $Sk_w$ obtained from the second term are excellent in the ISL and resemble the observed z-independent behaviour. Besides providing further confidence on the proposed theory, this result indicates that, since $K_{t,u}$ is directly proportional to $z$, $\partial \sigma_u^2/\partial z$ must overall scale as $\sim 1/z$, as predicted by the AEM. Hence, we argue that the AEM represents a reasonable approximation provided $B_u$ and $A_w$ are adjusted to accommodate for low $Re_{\tau}$ effects. As shown in Figure \ref{Fig:AEM_profile}, this is the case for both DNS and laboratory data.

For the DNS, appropriate values of $A_w$(=1.15-1.17) and $B_u$(=0.85-0.86) were estimated by fitting the AEM to the available data for all available $Re_{\tau}$. The constant $\kappa$=0.39 was assumed as reported in the literature \citep{marusic2013logarithmic,peruzzi2020scaling}. When inserting these choices of $A_w$ and $B_u$ from the DNS into equation \ref{eq:budgetwww6}, the computed $Sk_w=0.13$, which is close to reported values in Figure \ref{Fig:Skw_profile}c. The same approach was used for all laboratory studies. When combining all the runs together (wind tunnel and open channel experiments), ensemble-averaged $A_w=1.01\pm0.074$ and the ensemble-averaged $B_u=0.76\pm0.14$ were obtained across runs within an experiment and across experiments. These values result in an ensemble-averaged $Sk_w=0.18\pm0.07$ and agrees with the measurements reported in Figure \ref{Fig:Skw_profile}. 

This analysis and Figure \ref{Fig:Skw_profile} suggest that $Sk_w$ for DNS and experiments is higher than $0.12$, as estimated for $Re_{\tau}\to\infty$. This is probably because of deviations in the ISL variance statistics from their asymptotic constants in Equation \ref{eqn:law_of_wall}. The effects of such deviations on $Sk_w$ are however modest because, while low-$Re_{\tau}$ flows are characterized by values of  $A_w$ and $B_u$ that are significantly lower than their counterparts at $Re_{\tau}\to\infty$ (i.e. $A_w=1.33$, and $B_u=1.26$, see Table \ref{tab:table1}), equation \ref{eq:budgetwww6} indicates that $Sk_w$ is dictated by the ratio $B_u/A_w^3$, and hence the effect of such deviations are in good part compensated. 


Regarding the second feature, equation \ref{eq:budgetwww5} offers an explanation as to why conventional down-gradient closure models with eddy viscosity $K_{t}\propto\overline{q} l_m$ ($l_m$ is a 'master' mixing length) expressed in general index notation ($[x_1, x_2, x_3]=[x,y,z]$ and $[u'_1, u'_2, u'_3]=[u',v',w']$) as \citep{launder1975progress}
\begin{equation}
\label{eqn:www_gradww}
\overline{u_i'u_j'u_k'}=-K_{t} \left[\frac{\partial\overline{u_i'u_j'}}{\partial x_k}+\frac{\partial\overline{u_i'u_k'}}{\partial x_j}+\frac{\partial\overline{u_j'u_k'}}{\partial x_i}\right]
\end{equation}
spectacularly fail when $i=j=k=3$ and when $A_w$ is approximately constant in the ISL (as in the AEM). Yet, the derived equation here also offers a rectification based on the AEM.  This rectification accommodates the role of finite $\partial \sigma_u^2/\partial z$ on $\overline{w'^3}$ that cannot arise from equation \ref{eqn:www_gradww}.  While studies of the failure of gradient-diffusion models have a long tradition in turbulence research \citep{corrsin1975limitations}, the mode of failure these studies identify differ from the standard one offered here.  The classical failure of the so-called gradient diffusion phenomenology is the lack of accommodation of flux transport terms \citep{cava2006buoyancy,corrsin1975limitations}.  The origin of the failure of equation \ref{eqn:www_gradww} here is attributed to a return-to-isotropy mechanism (i.e. pressure-velocity destruction) that requires a model for the vertical transport of $K$, which is dominated by $\sigma_u^2$ \citep{lopez1999wall}.  To be clear, equation \ref{eq:budgetwww1} does maintain the turbulent flux transport term of $\overline{w'^3}$ through $\partial \overline{w'^4}/\partial z$, which is modeled using a quasi-normal approximation.  However, this term is not the main reason why equation \ref{eqn:www_gradww} fails. In fact, the final $Sk_w$ expression is independent of $a$ when $A_w$ is constant (as in AEM) irrespective of whether $a=0$ or not.  What was earlier assumed was $a$ not vary appreciably with $z$ within the ISL.

In conclusion, this letter demonstrates that, within the ISL of turbulent and adiabatic smooth wall flows, $Sk_w$ attain z-independent values that are predictable from well known turbulence constants relating to the AEM. This behavior is reported for a variety of different wall flows and is independent of variations in $Re_{\tau}$, hence universal and robust.

\bibliography{turbbib}
\begin{acknowledgments}
EB acknowledges Politecnico di Torino (Italy) for supporting the visit to Duke University. GK acknowledges support from the U.S. National Science Foundation (NSF-AGS-2028633) and the U.S. Department of Energy (DE-SC0022072). DV and CM acknowledge European Union’s Horizon 2020 research and innovation programme under the Marie Sklodowska-Curie grant agreement No 101022685 (SHIEELD).   DP acknowledges support from Fondo europeo di sviluppo regionale (FESR) for project Bacini Ecologicamente sostenibili e sicuri, concepiti per l'adattamento ai Cambiamenti ClimAtici (BECCA) in the context of Alpi Latine COoperazione TRAnsfrontaliera (ALCOTRA) and project Nord Ovest Digitale e Sostenibile - Digital innovation toward sustainable mountain (Nodes - 4). 
\end{acknowledgments}







\end{document}